\documentclass{article} 
\usepackage{iclr2023_conference,times}

\usepackage{amsmath,amsfonts,bm}

\def\eqref#1{equation~\ref{#1}}

\def\1{\bm{1}}

\DeclareMathAlphabet{\mathsfit}{\encodingdefault}{\sfdefault}{m}{sl}
\SetMathAlphabet{\mathsfit}{bold}{\encodingdefault}{\sfdefault}{bx}{n}

\usepackage{hyperref}
\usepackage{url}
\usepackage{braket}
\usepackage{mathtools}
\usepackage{subcaption}
\usepackage{siunitx}
\usepackage{enumitem}

\newcommand{\imodel}{\mathbf{I}}
\newcommand{\fmodel}[1]{ \mathcal{F} \left\{ \mathbb{E} \left[ #1 \right] \right\} }
\newcommand{\parbasic}[1]{\textbf{#1} \hspace{0.5mm}}

\newcommand*{\affaddr}[1]{\small #1}
\newcommand*{\affmark}[1][*]{\textsuperscript{#1}}
\newcommand*{\email}[1]{\small \texttt{#1}}

\title{Designing Nonlinear Photonic Crystals for High-Dimensional Quantum State Engineering}

\author{%
\parbox{\linewidth}{\raggedright 
            Eyal Rozenberg\affmark[1,*]~~~
            Aviv Karnieli\affmark[2]~~~
            Ofir Yesharim\affmark[3]~~~
            Joshua Foley-Comer\affmark[3]~~~
            Sivan Trajtenberg-Mills\affmark[4]~~~
            Sarika Mishra\affmark[5]~~~
            Shashi Prabhakar\affmark[5]~~~
            Ravindra Pratap\affmark[5]~~~
            Daniel Freedman\affmark[6]~~~
            Alex M. Bronstein\affmark[1]~~~ 
            and~~~ 
            Ady Arie\affmark[3]%
        }\smallskip \\
\affaddr{\affmark[1]Department of Computer Science, Technion, Haifa, Israel}\\
\affaddr{\affmark[2]School of Physics \& Astronomy, Tel Aviv University, Israel}\\
\affaddr{\affmark[3]}School of Electrical Engineering, Tel Aviv University, Israel\\
\affaddr{\affmark[4]Massachusetts Institute of Technology, Cambridge, MA, USA}\\
\affaddr{\affmark[5]}Quantum Technologies Laboratory, Physical Research Laboratory, Ahmedabad, India\\
\affaddr{\affmark[6]}Verily Research, Haifa, Israel\\
\smallskip
\email{\affmark[*]\href{mailto:eyal.rozenberg1@gmail.com}{eyal.rozenberg1@gmail.com}}
}

\iclrfinalcopy 
\begin{document}

\maketitle

\begin{abstract}
We propose a novel, physically-constrained and differentiable approach for the generation of D-dimensional qudit states via spontaneous parametric down-conversion (SPDC) in quantum optics. We circumvent any limitations imposed by the inherently stochastic nature of the physical process and incorporate a set of stochastic dynamical equations governing its evolution under the SPDC Hamiltonian. We demonstrate the effectiveness of our model through the design of structured nonlinear photonic crystals (NLPCs) and shaped pump beams; and show, theoretically and experimentally, how to generate maximally entangled states in the spatial degree of freedom. The learning of NLPC structures offers a promising new avenue for shaping and controlling arbitrary quantum states and enables all-optical coherent control of the generated states. We believe that this approach can readily be extended from bulky crystals to thin Metasurfaces and potentially applied to other quantum systems sharing a similar Hamiltonian structures, such as superfluids and superconductors.
\end{abstract}

\section{Introduction} \label{sec:introduction}
Quantum Optics \citep{scully1999quantum, garrison2008quantum} has proven to be an invaluable resource for the realization of quantum information systems \citep{ursin2007entanglement,gisin2007quantum,vallone2015experimental,chen2021integrated}. It is based on the transfer of data using single photons, where the information is encoded using a certain property of light (e.g., a photon’s polarization, color, or spatial shape). The unique quantum entanglement property can guarantee complete immunity to eavesdropping, using protocols such as \citet{PhysRevLett.67.661}. 
A key open question is how to design sources that can be used for quantum information protocols. A natural formulation is based on inverse problems \citep{tarantola2005inverse}, which aim at finding novel experimental setups that produce a desired physical observable. 
If we wish to employ learning-style optimization methods to solve such inverse problems, it is crucial to have a good physical model of the quantum process in question and integrate it into the algorithm itself \citep{choo2020fermionic,hermann2020deep, karniadakis2021physics, batzner20223}.
The model should ideally encompass the relevant conservation laws, physical principles, and phenomenological behaviors. Such physically-constrained models ensure convergence to physically realizable solutions, reduce the parameter search, improve the predictive accuracy of the model, and allow for faster training with improved generalization.

One of the most common processes used to produce entangled photon pairs is spontaneous parametric down conversion (SPDC), whereby a laser light beam illuminates a second order $\chi^{(2)}$ nonlinear photonic crystal (NLPC) \citep{SPDCreview2018}. The nonlinear coefficient of ferroelectric materials can be modulated by electric field poling in two out of the three crystal axes \citep{berger1998nonlinear, broderick2000hexagonally, ellenbogen2009nonlinear}. Recently, this capability has been extended to enable modulation in all three axes using focused laser beams  \citep{xu2018three,wei2018experimental, arie2021storing}, which introduces additional degrees of freedom for tailoring the quantum state. Another possibility to control the SPDC process is by shaping the input pump beam.
The laser beam has a myriad of photons and occasionally one of the photons will spontaneously decay inside the nonlinear crystal and produce a highly correlated photon pair. This pair can be entangled in many possible degrees of freedom. In this work, we focus on the ability to entangle the photon pair in the spatial degree of freedom, whereby different free space orthogonal modes, e.g. of the Hermite-Gaussian or Laguerre-Gausian basis, can be used. The high dimensionality of these generated states increases the bandwidth of quantum information \citep{brandt2020high} and can improve the security of quantum key distribution protocols \citep{krenn2015twisted,sit2017high,sit2018quantum}. 
We employ a machine learning algorithm to find the conditions that will generate the photon pair with the desired entanglement in the spatial domain, using tailored nonlinear interactions in the SPDC process. We validate our model against current and previous experimental results. We also show how a generated high-dimensional maximally entangled quantum state can be coherently controlled by altering the pump shape – a feature that can find applications in qudit-based quantum key distribution and quantum information protocols that work at high switching rates. The entire algorithm is released as open source \citep{jax-spdc_inv}. We encourage the reader to refer to our recently published paper for more details about the algorithm \citep{rozenberg2022inverse}.

\section{Methodology} \label{sec:methodology}

\parbasic{SPDC Forward Model} We consider SPDC in a bulk nonlinear crystal of uniform refractive index and spatially-varying second order nonlinearity, $\chi^{(2)}$, and show how to make an inherently stochastic description of SPDC fully differentiable. The SPDC forward model captures the interaction properties, such as diffraction, space-dependent nonlinear coupling, vacuum fluctuations and non-perturbative effects; and respects conservation laws, such as momentum and energy. Furthermore, the model makes it possible to accurately compute the correlations between the two photons created in the SPDC process, with the interaction properties used as parameters on which learning can be performed \citep{rozenberg2022inverse}. The dynamics are prescribed by the Heisenberg equations of motion: $i\hbar \partial_t \hat{E} = [\hat{E},\hat{H}_{\mathrm{SPDC}}]$, for the field operators $\hat{E}$ evolving under the SPDC Hamiltonian $\hat{H}_{\mathrm{SPDC}}$, where $\hbar$ is the reduced Planck's constant;
which can be described by two pairs of c-number coupled wave equations along the interaction medium: 
\begin{equation}
\begin{split}
    i\frac{\partial E_{i}^{out}}{\partial \zeta}  = -\frac{\nabla^2_\perp}{2k_i} E_{i}^{out}+\kappa_ie^{-i\Delta k \zeta}(E_{s}^{vac})^* \quad \quad  i\frac{\partial E_{i}^{vac}}{\partial \zeta}  =   -\frac{\nabla^2_\perp}{2k_i}E_{i}^{vac}+\kappa_ie^{-i\Delta k \zeta}(E_{s}^{out})^* \\
    i\frac{\partial E_{s}^{out}}{\partial \zeta}  = -\frac{\nabla^2_\perp}{2k_s}E_{s}^{out}+\kappa_se^{-i\Delta k \zeta}(E_{i}^{vac})^* \quad \quad  i\frac{\partial E_{s}^{vac}}{\partial \zeta}  = -\frac{\nabla^2_\perp}{2k_s}E_{s}^{vac}+\kappa_se^{-i\Delta k \zeta}(E_{i}^{out})^*
\label{eq:waveeq}
\end{split}
\end{equation}
where $\zeta=z$ is the coordinate along the direction of propagation. In the above equation: $E_{j}^{out},E_{j}^{vac}$ ($j=i,s$ for the idler and signal fields respectively) are the matrix elements representing the ``output'' and ``vacuum'' field amplitudes.
$\nabla^2_\perp$ is the transverse Laplacian operator; $k_j$ is the wavenumber; $\kappa_{j} (\textbf{r}, \zeta)=\frac{\omega_j^2}{c^2 k_j} \chi^{(2)} (\textbf{r}, \zeta) \mathcal{E}_{p}(\mathbf{r})$ is the nonlinear-coupling coefficient, where $\textbf{r}=(x,y)$ is a position on the transverse plane; $\chi^{(2)}(\textbf{r}, \zeta)$ stands for the (spatially varying) second-order susceptibility and $\mathcal{E}_{p}(\mathbf{r})$ is the (spatially varying) pump field envelope; $c$ is the speed of light in vacuum; and $\Delta k=k_p-k_s-k_i$ is the phase mismatch. The quantum vacuum noise is emulated by initializing a large number of instances of Gaussian noise in both the idler and signal amplitudes at $z=0$.

\parbasic{Solving the Forward Problem} To solve Eq. \ref{eq:waveeq} we integrate the fields along the direction of propagation and solve the coupled wave equations for the large ensemble of quantum vacuum realizations in parallel. We use a LISTA-like time-unfolded version \citep{gregor2010learning} of the Split-Step Fourier method \citep{stoffa1990split, agrawal2001applications} to solve for the propagation along the crystal.  Note that this technique is also relevant for many other inverse problems in optics and quantum mechanics, as it combines diffraction, or more generally propagation in space, to solve nonlinear partial differential equations like the nonlinear Schr\"{o}dinger equation. We then derive the second-order statistics to and describe the full quantum state generated by the SPDC process by quantum state tomography (QST) \citep{thew2002qudit, agnew2011tomography}.

\parbasic{Solving the Inverse Problem} This strategy facilitates differentiation back through the model and enables application of powerful optimization methods for learning its physical parameters, thereby overcoming issues related to the fundamentally stochastic nature of the model. We rewrite our forward model as
\begin{equation}
\label{eq:forward_model}
\mathbb{O} = \fmodel{\mathcal{P}(\Lambda)},
\end{equation}
where $\mathbb{O}$ is the set of observables of interest, such as coincidence rate count $G^{(2)}$ and the density matrix of the bi-photon quantum state $\rho$ (discussed in more detail in Appendix \ref{sec:observables}); $\mathcal{P}(\Lambda)$ denotes the solution of Eq. \ref{eq:waveeq} for the set of parameters $\Lambda$ and a particular realization of the vacuum noise, followed by projection of the output and noise fields onto a desired orthonormal basis; $\mathbb{E}$ denotes the expectation over the vacuum noise; and the operator $\mathcal{F}$ computes the first-order correlations which yield the desired observable. Given a desired observable-set, $\mathbb{O}_d$, describing the quantum state or any related features, our goal is to find the unknown physical parameters, $\Lambda$, that characterize the system. We take a parameterized approach to solving the inverse problem, i.e. $\Lambda = \Lambda(\theta)$, and solve the inverse problem by solving the optimization problem
\begin{equation}
\label{eq:optimizer}
\theta^* = \min_\theta \mathcal{D}\big( \fmodel{\mathcal{P}(\Lambda(\theta))}, \mathbb{O}_d \big)
\end{equation}
In the above, $\mathcal{D}(\cdot, \cdot)$ is a discrepancy measure between two sets of observables; for example, in the case where we are measuring the discrepancy between two density matrices, we may take $\mathcal{D}$ to be the Trace Distance \citep{rana2016trace}. In Eq. \ref{eq:optimizer}, by minimizing the discrepancy the model produces the properties of the optical system that would produce a result as close as possible to the desired quantum state. The inverse model is then given by
\begin{equation}
\label{eq:model}
\imodel(\mathbb{O}_d) = \Lambda(\theta^*),
\end{equation}

where $\imodel(\cdot)$ is our inverse solver. In order to solve the optimization problem in Eq. \ref{eq:optimizer}, an approach based on gradient descent may be employed.  The key is that the forward model of Eq. \ref{eq:waveeq}, while quite complicated, can be expressed in such a way that it is fully differentiable.  As a result, any library which can auto-differentiate a system may be used to compute the relevant gradients, thereby allowing for the solution to the optimization problem in Eq. \ref{eq:optimizer}.  In practice, we use JAX. We may learn any physical parameters $\Lambda$ of the interaction, e.g. wavelength, temperature profile, poling period, poling profile, etc. In this work, the 3D NLPC structure, $\chi^{(2)}(\textbf{r}, \zeta)$, and pump beam profile, $\mathcal{E}_{p}(\textbf{r})$, are the unknown physical parameters we seek to learn, that is $\Lambda = (\mathcal{E}_{p}(\cdot), \chi^{(2)}(\cdot))$. We parameterize the 2D/3D NLPC structure and pump beam profile by the multi-dimensional parameters $\theta_\mathcal{E}$ and $\theta_\chi$, respectively, such that $\Lambda(\theta) = (\mathcal{E}_{p}(\cdot; \theta_\mathcal{E}), \chi^{(2)}(\cdot; \theta_\chi))$. We discuss in more detail how this parameterization is performed in Appendix \ref{sec:interaction_params}.


\section{Results} \label{sec:results}
We use our algorithm to solve the inverse design problem and extract the optimal NLPC structures and the complex pump beam structures (according to Eq. \ref{eq:parameterized}) for generating desired second-order quantum correlations or density matrices. The training phase takes about one hour on 4 nvidia t4 16gb gpus, for all configurations involving 1mm-long NLPCs.

\parbasic{Model Validation}
Our model was able to recover experimental results reported by \citet{PhysRevA.98.060301} and reproduce the coincidence rate counts, in the Laguerre-Gauss basis, for a qutrit state (Fig. \ref{fig:Kovlakov_PRA}a) and ququint state (Fig. \ref{fig:Kovlakov_PRA}b), and the density matrix of the qutrit state (Fig. \ref{fig:Kovlakov_PRA}c). Results are generated by a shaped pump field and measurement are performed in the LG basis. Additional comparisons in tabular form are presented in Appendix \ref{sup:comparison}. We follow another result reported by \citet{kovlakov2017spatial} et~al. and let our algorithm learn the optimal pump waist size for generating a pure HG spatial Bell state between structured SPDC photon pairs. Fig. \ref{fig:Kovlakov_PRL} shows the convergence of our learning algorithm towards the optimal pump waist, $w_p=\sqrt{L/k_p}$ \citep{kovlakov2017spatial}. As the learning process progresses, the discrepancy measure $\mathcal{D}(\cdot, \cdot)$ in Eq. \ref{eq:optimizer} decreases until the model reaches convergence. Concomitantly, the size of the pump waist converges to the desired value \citep{kovlakov2017spatial} and a clear Bell state, $(\ket{0,1}+\exp(i\phi)\ket{1,0})/\sqrt{2}$, is generated.

\parbasic{Experiments} We now experimentally demonstrate the effectiveness of our model in the discovery of quantum states. Our first experimental setup measures the orbital angular momentum (OAM) correlation between two photons by the design of pump structure.  The experimental setup and procedure are detailed in Appendix \ref{sup:pump_shaping} and in Fig. \ref{fig:exp}. First, we show the correctness of our forward model \ref{eq:forward_model} and examine the effect of the pump beam waist and the coupling efficiency on the OAM spectrum. Figs. \ref{fig:exp_oam_spect1} \& \ref{fig:exp_oam_spect2} present good correspondence between numerical and experimental results. Next, we use the model to learn the pump structure in order to obtain the second order quantum correlation corresponding to a desired qutrit quantum state, $\ket{\psi}=(\ket{-1,1}+\exp(i\phi_1)\ket{0,0}+\exp(i\phi_2)\ket{1,-1})/\sqrt{3}$. We then use the learned pump structure to reproduce the quantum state experimentally; in Fig. \ref{fig:exp_qutrit}, the experimental outcome resulting from the use of the learned pump reproduces the desired coincidence rate counts. The second experimental setup demonstrates the correctness of our forward model by shaping the spatial quantum correlations of entangled photon pairs in 2D patterned NLPC \citep{ofir_holo}. The experimental setup is detailed in Appendix \ref{sup:crystal_shaping} and in Fig. \ref{fig:exp2}. Different NLPC structures were used to shape the quantum correlations between the down converted photons. We find good agreement between experiments and numerical simulation.

\parbasic{Theoretical Extensions}
We demonstrate that the quantum state of SPDC photons and their correlations can be all-optically controlled, by first learning the 3D crystal structure with a given pump mode, and then changing the initial pump mode in inference phase. This active optical control has the advantage of altering the quantum state in a non-trivial manner, while retaining its purity. In this case, the discrepancy measure in Eq. \ref{eq:optimizer} is taken as a weighted ensemble of the Kullback-Leibler divergence and the L1 norm. Fig. \ref{fig:lg1} depicts the results of this theoretical experiment for the generation of desired coincidence rate counts of a maximally-entangled two-photon qubit state $\ket{\psi}=(\ket{1,-1}+\exp(i\phi)\ket{-1,1})/\sqrt{2}$ and ququart state $\ket{\psi}=(\ket{-2,1}+\exp(i\phi_1)\ket{0,-1}+\exp(i\phi_2)\ket{-1,0}+\exp(i\phi_3)\ket{1,-2})/\sqrt{4}$, where we project the generated photons on the LG modes with the integer quantum numbers $l,p$, standing for the azimuthal and radial numbers, respectively. As we alter the initial pump mode, the new correlations differ significantly from those obtained in the original design, while still corresponding to maximally-entangled states with high SNR.

In order to resolve a specific two-photon quantum state generated by the tailored SPDC process, a coincidence measurement will not suffice. Thus, we emulate QST and integrate it into our learning stage for evaluating the corresponding density matrix, as detailed in Appendix \ref{sec:observables}. Here we focus on the subspace spanned by $\lbrace\ket{-1}, \ket{0}, \ket{1} \rbrace\otimes\lbrace\ket{-1}, \ket{0}, \ket{1}\rbrace$, giving a 9-by-9 dimensional density matrix. The density matrix is used as an observable while $\mathcal{D}(\cdot, \cdot)$ is taken to be the Trace Distance  (Eq. \ref{eq:optimizer}). Our algorithm simultaneously extracts the optimal 3D NLPC structures and the pump beam profiles, for generating the desired quantum states. Fig. \ref{fig:rho1}a depicts the results for the maximally-entangled state $\ket{\psi}=(\ket{1,-1} + \ket{-1,1})/\sqrt{2}$ (corresponding to the coincidence rate shown in Fig. \ref{fig:lg1}a(i)), while Fig. \ref{fig:rho1}b depicts the results for the maximally-entangled state $\ket{\psi}=(\ket{1,-1} + \ket{0,0} +\ket{-1,1})/\sqrt{3}$. The simultaneous learning makes higher-order radial LG modes possible.  This is responsible for removing the two-photon Gaussian mode $\ket{00}$ in the first learned state (Figs. \ref{fig:rho1}a(i) and \ref{fig:lg1}a(ii)) through destructive interference, which is impossible when only using Gaussian pump beams. Importantly, the generated quantum two-photon states are sensitive to the relative phase between the modes constituting the pump profile and the learned  crystal structure; which implies that the active all-optical control over the coincidence rate counts 
also allows for quantum coherent control over the generated photon qudits. In Fig. \ref{fig:rho2}, we again learn a 3D crystal structure with a fixed pump profile, but now consisting of a given superposition of LG modes. By changing the relative phase between the LG modes, we expect that the off-diagonal terms in the density matrix will change accordingly. This corresponds experimentally to a rotation of the $\mathrm{HG}_{10}$ mode.

\parbasic{Robustness} To mimic crystal fabrication imperfections we deliberately add errors to the crystal structure to corrupt the generated coincidence rate counts of the maximally-entangled two-photon qubit.  In Appendix \ref{sec:robustness} and Fig. \ref{fig:suppTolerance}, we show how with a slight variation in a different parameter of the system (pump waist), we can nearly recover the original system results. In particular, the imperfections cause the model to diverge from the optimum for generating the desired quantum state; however, since the model was very close to a global optimum, a slight variation in a different parameter of the system (pump waist) allows the system to revert.


\section{Conclusions and Directions for Future Work} \label{sec:conclusion}
We have shown how machine learning algorithms can be used to solve an open problem in quantum optics.
Our model achieves new and highly desired quantum states through the design of   second order $\chi^{(2)}$ nonlinear photonic crystals and shaped pump beams. We further show the diverse functionality available by use of a single crystal structure pumped with different optical modes for the coherent control over the quantum state via the modification of the pump beam. We believe that this approach can readily be extended from bulky crystals towards the realm of thin Metasurfaces \citep{santiago2022resonant}; and may be adapted to other quantum systems sharing a similar Hamiltonian structure, such as superfluids and superconductors \citep{coleman2015introduction}, or for other optical systems, such as nonlinear waveguides and resonators \citep{QiLi+2020+1287+1320}. The model can be further extended to control other degrees of freedom of quantum light, such as the frequency of the signal and idler, and can be also studied in the case of high parametric gain. Finally, whereas here we studied second order nonlinear processes, our model can be easily extended to study third order nonlinear effects  e.g. spontaneous four wave mixing \citep{sharping2006generation}.
\subsubsection*{Acknowledgments}
Israel Science Foundation, Israel Ministry of Science.

\bibliography{iclr2023_conference}
\bibliographystyle{iclr2023_conference}

\appendix
\section{Observables}\label{sec:observables}
The set of desired observables describing the generated quantum state is given by the coincidence rate count, $G^{(2)}$, and density matrix of the bi-photon quantum state, $\rho$, such that in general $\mathbb{O}_d = (G^{(2)}_d, {\rho}_d)$. Their evaluation is achieved by first solving Eq. \ref{eq:waveeq} over a large number of independent realizations of the vacuum noise, projecting the output and noise fields onto a desired orthonormal basis of optical modes, and then taking the ensemble average to obtain first-order correlations \citep{brambilla2004simultaneous, trajtenberg2020simulating, rozenberg2021inverse, rozenberg2022inverse, Rozenberg2022}, which (for the signal) is given by $G^{(1)}(q_s,q'_s)=\braket{\psi|a^{\dagger}_{q_s}a_{q'_s}|\psi}$. Here, $\ket{\psi}$ denotes the quantum state, $a$ ($a^{\dagger}$) denotes the photon annihilation (creation) operator, and $q_s$ denotes any quantum number of the signal photon, for example, LG modes, HG modes, etc. Second-order correlations are derived using the fact that the quantum state of SPDC, the squeezed vacuum state \citep{wu1986generation}, belongs to the family of Gaussian states, for which all higher-order correlations can be obtained from the first-order ones \citep{gardiner2004quantum, Rozenberg2022}. The coincidence rate is given by the second-order quantum correlation function, which determines the probability of finding an idler photon in mode $q_i$ and a signal photon in mode $q_s$

\begin{equation}
G^{(2)}(q_i,q_s,q_s,q_i)=\braket{\psi|a^{\dagger}_{q_i}a^{\dagger}_{q_s}a_{q_s}a_{q_i}|\psi}
\label{eq:G2}
\end{equation}

To extract the optimal model parameters that generate the desired quantum correlations over a given basis, we solve the optimization problem in Eq. \ref{eq:optimizer}. Here, $\mathcal{D}(\cdot, \cdot)$ is taken as a typical measure of discrepancy between two probability distributions. For example, we may use the Kullback-Leibler divergence \citep{georgiou2003kullback}, the L1 norm \citep{gine2003bm}, or an ensemble of both.

To obtain the full quantum state generated by the SPDC process, we use quantum state tomography (QST) \citep{thew2002qudit, agnew2011tomography, toninelli2019concepts}. Eq. \ref{eq:G2} allows for the calculation of any coincidence measurement performed on the system, on any basis of our choice. Since the process of QST involves a sequence of projective coincidence measurements on different bases, we can readily reconstruct  the density matrix, $\rho$, of the entangled two-qudit state, through a series of linear operations. Here, naturally, $\mathcal{D}(\cdot, \cdot)$ (in Eq. \ref{eq:optimizer}) is taken to be the Trace Distance \citep{rana2016trace} -- a metric on the space of density matrices that measures the distinguishability between two states.

The tomographic reconstruction is performed using the correlation data collected from the projections of the simulated bi-photon state onto orthogonal as well as mutually unbiased bases (MUBs) \citep{toninelli2019concepts, agnew2011tomography}. The density matrix of the bi-photon system can be written as
\begin{equation}
    \rho  = \frac{1}{d^{2}}
    \sum_{m,n=0}^{d^{2}-1}\rho_{mn}\\
    \sigma_{m}\otimes\sigma_{n}
\end{equation}
where $\sigma_{m}$ are the set of generators that span the $d$-dimensional tomography space (for example, Pauli and Gell-Mann matrices for $d=2$ and $3$, respectively). The expansion coefficients $\rho_{mn}$ are found via 
\begin{equation}
    \rho_{mn}  =
    \sum_{i,j=0}^{d-1}\\
    a_{m}^{i}a_{n}^{j}\braket{\lambda_{m}^{i}\lambda_{n}^{j}|\\
    \rho|\lambda_{m}^{i}\lambda_{n}^{j}}
\end{equation}
with $a_{m}^{i}$ and $|\lambda_{m}^{i}\rangle$ denoting the $i^{th}$ eigenvalue and eigenstate of $\sigma_{m}$, respectively \citep{toninelli2019concepts}. The required projections inside the sum function are found in a similar manner to Eq. \ref{eq:G2}, with the pure basis states replaced by the MUBs, when necessary.

\section{Interaction Parameters} \label{sec:interaction_params}
The parameters we learn can be as general as we want, subject to technological and physical restrictions. To decrease the dimensionality of learned parameters in order to ensure smoother convergence of the inverse problem's solution, the continuous functions of the NLPC structures are represented using a finite set of unknowns. One way to do this is through expansion in set basis functions that are mutually orthogonal, which may also change as a function of the propagation coordinate, $\zeta$; the parameters $\theta$ then include the coefficients of the expansion. Examples include the Hermite-Gauss (HG) and Laguerre-Gauss (LG) bases, though many other possibilities exist. These basis functions are often scaled according to a transverse length, which for light beams is usually referred to as the waist size, a term which we adopt hereafter for all basis functions. Learning the waist sizes of each of the basis functions individually adds further degrees of freedom to our model. The exact role of the parameters can be seen by formally writing the NLPC structure and the pump profile as a linear combination of the basis functions:
\begin{align}
    \chi^{(2)}(\textbf{r}, \zeta; \theta_\chi) & = \sum_{n=1}^{N_\chi}{\alpha_\chi^n \Phi_\chi^n(\textbf{r}, \zeta; w_\chi^n)} & \quad \theta_\chi = \left\{ \left(\alpha_\chi^n, w_\chi^n \right) \right\}_{n=1}^{N_\chi} \notag \\
    \mathcal{E}_{p}(\textbf{r}; \theta_\mathcal{E}) & = \sum_{n=1}^{N_\mathcal{E}}{\alpha_\mathcal{E}^n \Phi_\mathcal{E}^n(\textbf{r}; w_\mathcal{E}^n)} & \quad \theta_\mathcal{E} = \left\{ \left(\alpha_\mathcal{E}^n, w_\mathcal{E}^n \right) \right\}_{n=1}^{N_\mathcal{E}}
\label{eq:parameterized}
\end{align}
where $\alpha_\chi^n, \alpha_\mathcal{E}^n$ are the learned basis coefficients; $w_\chi^n, w_\mathcal{E}^n$ are the learned basis function waist sizes; and $\Phi_\chi^n, \Phi_\mathcal{E}^n$ are the basis functions. Here, the basis function index $n$ sums over both transverse modal numbers, for example the orbital angular momentum (OAM) $l$- and radial $p$-indices for LG modes.

\section{Validation -- data comparison} \label{sup:comparison}
\begin{figure*}[]
\centering
\begin{tabular}{c}
    \includegraphics[width=0.95\linewidth]{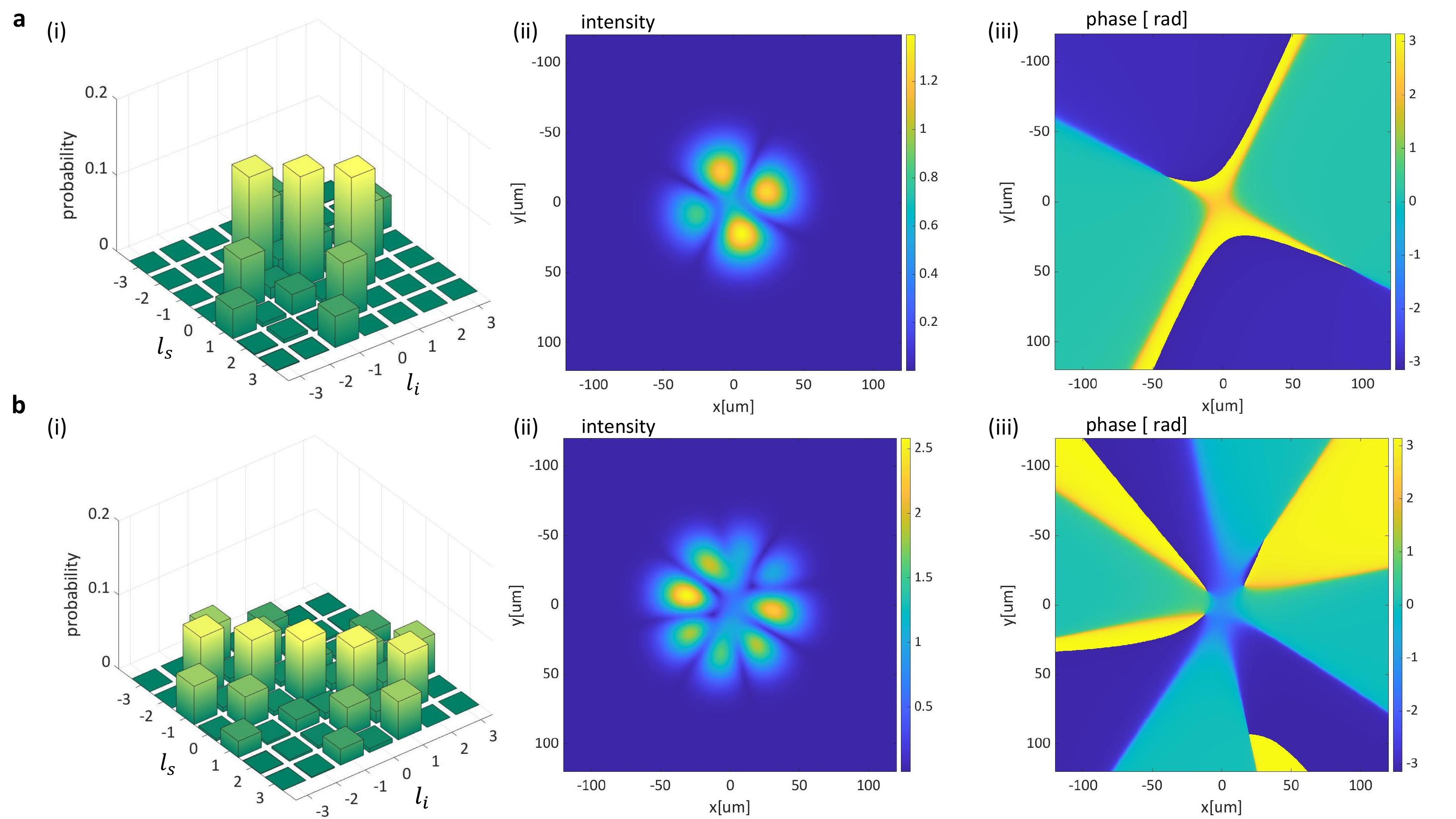}  \\
    
    \includegraphics[trim={0 9cm 0 0},clip,width=0.95\linewidth]{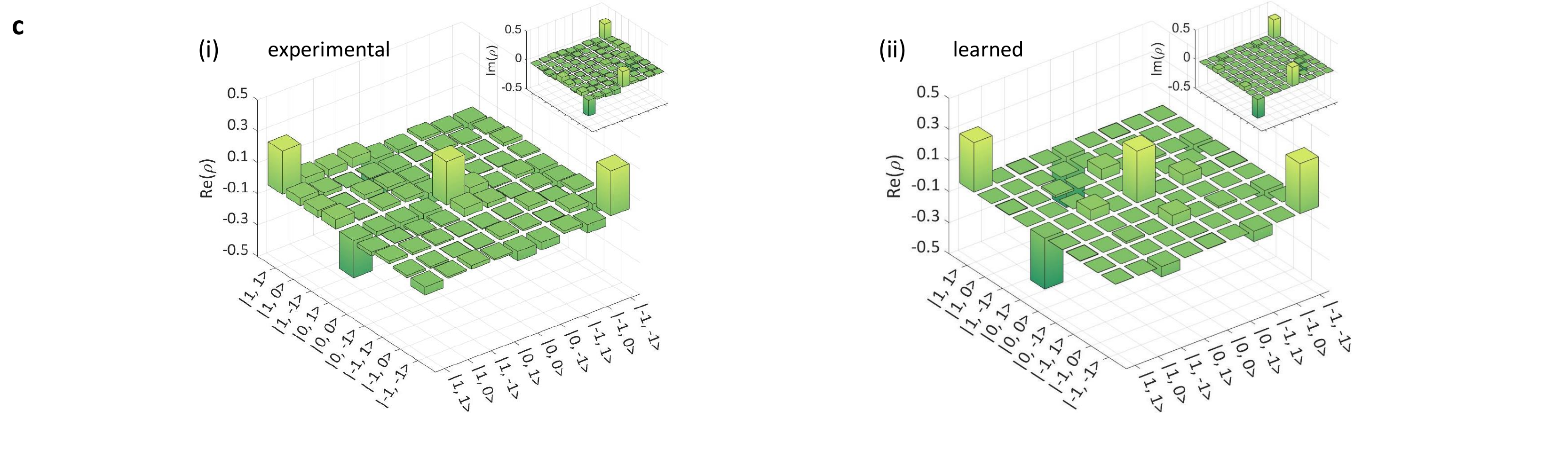} 
\end{tabular}
 \caption{ \small Model validation against experimental results reported by Kovlakov et al. \citep{PhysRevA.98.060301}. \textbf{a} LG qutrit state: model generated coincidence rate counts (i), and the corresponding pump intensity (ii) and phase (iii). \textbf{b} LG ququint state: model generated coincidence rate counts (i), and the corresponding pump intensity (ii) and phase (iii). \textbf{c} LG qutrit density matrix: experimental result \citep{PhysRevA.98.060301} (i) and model generated result (ii). $l_{s/i}$ are the LG modes' azimuthal integer quantum numbers for the signal/idler, respectively.}
 \label{fig:Kovlakov_PRA}
\end{figure*}

We compare between experimental setup reported by \citet{PhysRevA.98.060301} and our model's generated pump parameters used to recover the experimental results, as in Fig. \ref{fig:Kovlakov_PRA}. This provides further verification of the correctness of our model in the context of producing physically-constrained results. For the pump field coefficient amplitudes, we obtain mean squared errors (MSEs) of \num{1.59e-2} and \num{1.92e-2}, respectively (Table \ref{table:exp_qutrit_ququint_pumpCoeffs}). In the coincidence rate learning examples, the pump field propagates through a uniform crystal. As a result of angular momentum conservation, there is no interference between different pump field LG modes. This means that the phase of each pump field mode is a degree of freedom and the same coincidence rate can be achieved for infinitely many combinations of such different phases. For this reason, the MSE for the pump field modes was calculated with regards to the pump field amplitude alone.  Analyzing the pump field mode amplitudes (Table \ref{table:exp_qutrit_ququint_pumpCoeffs}), a very high level of symmetry is seen between the LG modes with opposite sign in the case of the model learned pump. This is the reason for the observed symmetry seen in the coincidence counts in Fig. \ref{fig:Kovlakov_PRA}a(i) and \ref{fig:Kovlakov_PRA}b(i). On the other hand, the experimental pump amplitudes do not exhibit such symmetry, even though the measured coincidences are symmetric with respect to changing OAM signs. This can further imply a possible asymmetry in the experimental setup, either in pump field preparation, mode projection, fiber coupling, or photodetection.

\begin{table}[h!]
\setlength{\arrayrulewidth}{0.5mm}
\setlength{\tabcolsep}{16pt}
\renewcommand{\arraystretch}{1}
\begin{tabular}{ |p{1cm}|p{1.8cm}|p{1.8cm}|p{1.8cm}|p{1.8cm}| }
\hline
Pump Mode& \multicolumn{2}{|c|}{Coefficients}&\multicolumn{2}{|c|}{Amplitudes} \\
\hline
& Experimental &Learned &Experimental &Learned\\
\hline
\multicolumn{5}{|c|}{Qutrit Coincidence Counts, Fig. \ref{fig:Kovlakov_PRA}a(ii)-(iii)} \\
\hline
$\mathrm{LG_{0-2}}$ &0.76-0.11i &0.46+0.45i &0.77 &0.64 \\
$\mathrm{LG_{00}}$  &-0.12+0.15i &-0.24+0.23i &0.19 &0.33 \\
$\mathrm{LG_{02}}$ &0.30-0.53i &0.59-0.30i &0.61 &0.66\\
\hline
\multicolumn{5}{|c|}{Ququint Coincidence Counts, Fig. \ref{fig:Kovlakov_PRA}b(ii)-(iii)} \\
\hline
$\mathrm{LG_{0-4}}$ &0.25-0.73i &0.41-0.40i &0.77 &0.57 \\
$\mathrm{LG_{0-2}}$ &0.19-0.10i &0.24+0.26i &0.21 &0.35 \\
$\mathrm{LG_{00}}$ &-0.07+0.11i &0.10-0.15i &0.13 &0.18\\
$\mathrm{LG_{02}}$ &0.14-0.14i &0.24+0.27i &0.20 &0.36 \\
$\mathrm{LG_{04}}$  &-0.54+0.09i &0.40+0.39i &0.55 &0.56 \\
\hline
\end{tabular}
\caption{Tabular comparison of the experimental coefficients \citep{PhysRevA.98.060301} and the pump field coefficients and pump field coefficient amplitudes learned by the model in Fig. \ref{fig:Kovlakov_PRA}a and \ref{fig:Kovlakov_PRA}b (ii)-(iii)}
\label{table:exp_qutrit_ququint_pumpCoeffs}
\end{table}

\section{Experimental Setups} \label{sup:experimental_setups}
\subsection{Pump Shaping}\label{sup:pump_shaping}
\begin{figure}[h]
    \centering
    \includegraphics[width=0.75\linewidth]{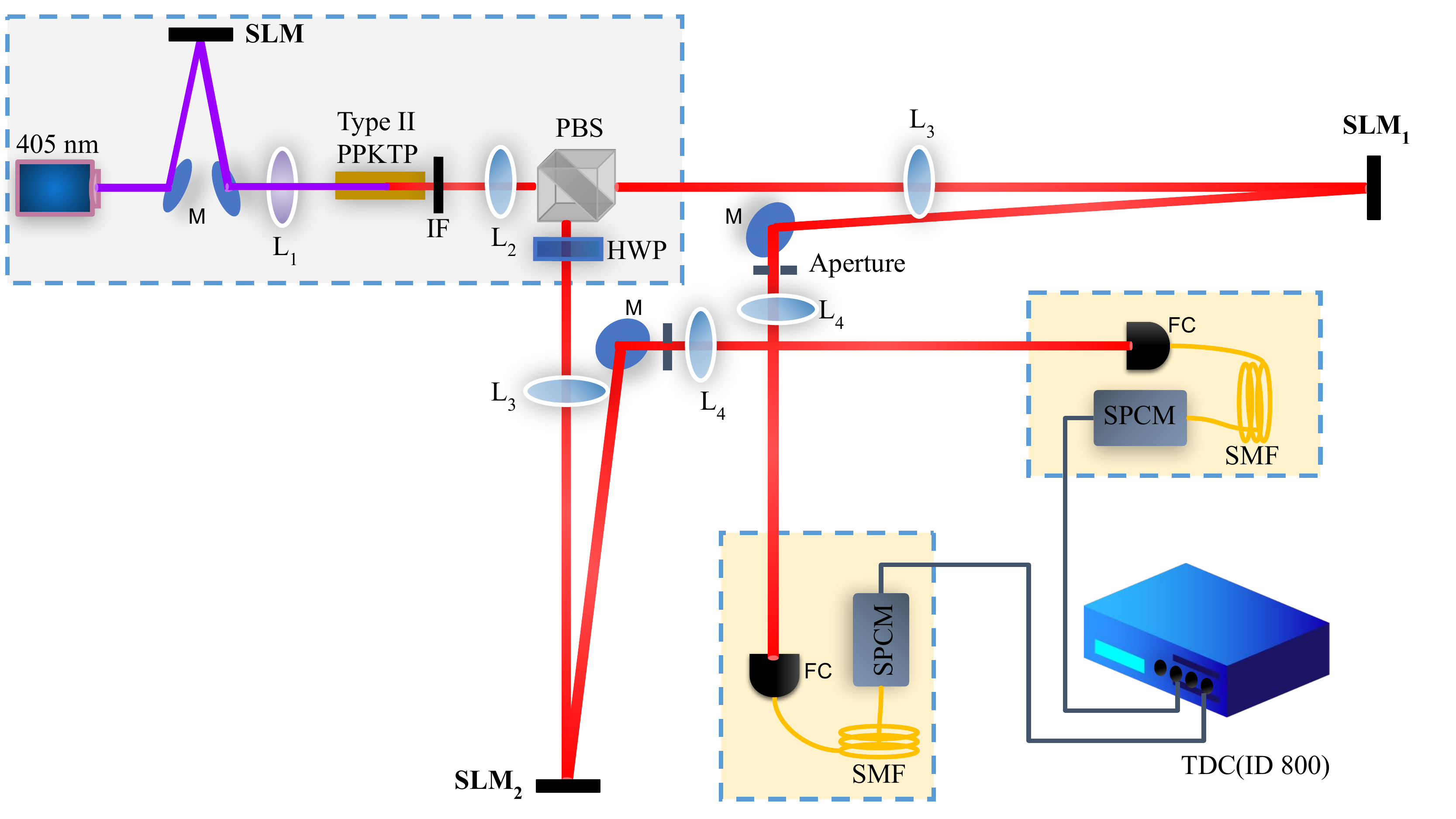}
    \caption{ \small Experimental setup for measuring the OAM correlation in SPDC. SLM is used to change the spatial mode of Pump. Lens $L_1$ is used to focus the pump at the center of Type-II PPKTP crystal. combination of lens $L_2$ and $L_3$ is used to image the crystal plane on SLM$_1$ and SLM$_2$. SLMs plane are imaged onto the SMF for projective measurements.}
    \label{fig:exp}
\end{figure}

The experimental setup to measure the OAM correlation between two photon, signal and idler, is shown in Figure \ref{fig:exp}. A 250 mW ultra violet (UV) 405 nm diode laser with a spectral band-width of 2 nm is used to pump a 30 mm long second order nonlinear Type-II periodically-poled KTP (PPKTP) crystal with a period $\Lambda=10~\mu$m. Spatial light modulator (SLM) is used to change the spatial profile of pump. Lens $L_1$ is used to focus the pump at the center of the crystal to generate entangled photon pairs, signal and idler of wavelength 810 nm each. A interference filter (IF) (810$\pm$5)nm is used to block the pump beam after the crystal and pass down the signal and idler of wavelength 810 nm. Both the photons, signal and idler, split into two different directions after passing through the polarizing beam splitter (PBS). The down-converted photons, signal and idler, are then imaged to SLM$_{1}$ and SLM$_{2}$, using lens $L_2$ and $L_3$ (4f-imaging). SLM$_{1}$ and SLM$_{2}$ are used to perform projective measurement by projecting the signal and idler to a conjugate LG mode so that resultant output will become Gaussian. We selected first order diffraction of the output of each SLM. To measure the projected photon, the SLM plane is again imaged to the fiber couplers (FC) using lens $L_4$ and the aspheric lens attached with the fiber coupler ($f$=4.6 mm). The fiber couplers are attached to the single mode fibers (SMF) each having a mode field diameter of $5\pm0.5~\mu$m which are then connected to the single photon counting module having a time jitter of 350 ps. Both the SPCMs are connected to the coincidence counter (time resolution 81 ps) to measure number of correlated photon pairs.

\subsection{Crystal Shaping}\label{sup:crystal_shaping}
\begin{figure}[h]
    \centering
    \includegraphics[width=0.75\linewidth]{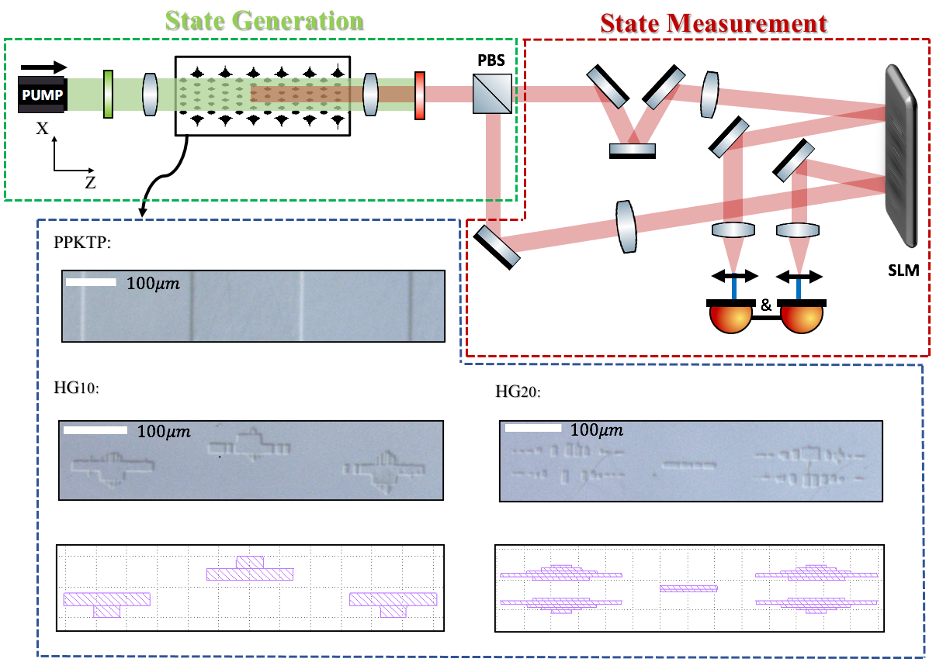}
    \caption{ \small Experimental setup. (In green \& red) A continuous wave (CW) 532.25 nm pump is focused to the patterned KTP crystal. The pump beam is filtered, and the photon pairs are split using a polarizing beam splitter (PBS). The $H$ and $V$ polarized photons are then sent to two halves of an SLM, after rotation of the $V$-polarized photon to an $H$-polarized photon (not shown in the figure). Photons from the first diffraction order of the SLM were coupled to two single mode fibers followed by two “Single Quantum” super-conducting nanowire single photon detectors (SNSPDs) for coincidence counting, with a coincidence window of 2.5 ns. (In blue, top rows)  Microscopic pictures of the top surface of the fabricated crystals after selective etching that reveals the nonlinear modulation pattern in the $\mathrm{HG_{10\\20}}$ case and in a regular PPKTP case. (In blue, bottom row) Original design of the nonlinear crystal. Left – $\mathrm{HG_{10}}$.  Right – $\mathrm{HG_{20}}$. (In courtesy of \citet{ofir_holo})}
    \label{fig:exp2}
\end{figure}

The experimental setup depicted in Fig. \ref{fig:exp2} measures the quantum correlations between different spatial mode profiles of the signal and idler photons generated using quantum NLPC. The quantum light source is based on degenerate Type-II SPDC process that generates two orthogonally polarized ($H$, $V$) 1064.5 nm photons. The photon pairs were split by a PBS and relay-imaged onto two halves of a rectangular phase only SLM. The SLM spatially matches between the shapes of the impinging photons and the shapes of the two single mode fibers that coupled the light to the detectors; thus, it acts as measurement tool for specific transverse spatial modes. By realizing a patterned blazed grating on the SLM, both the amplitude and phase of the photons incident on the SLM are manipulated. By changing the SLM's patterns and measuring the coincidence rate, the coefficients of the decomposed bi-photon state are mapped and verified the corresponding generated bi-photon quantum state.

\section{Effects of crystal imperfections} \label{sec:robustness}
\begin{figure*}[]
\centering
\begin{tabular}{c}
    \includegraphics[width=0.95\linewidth]{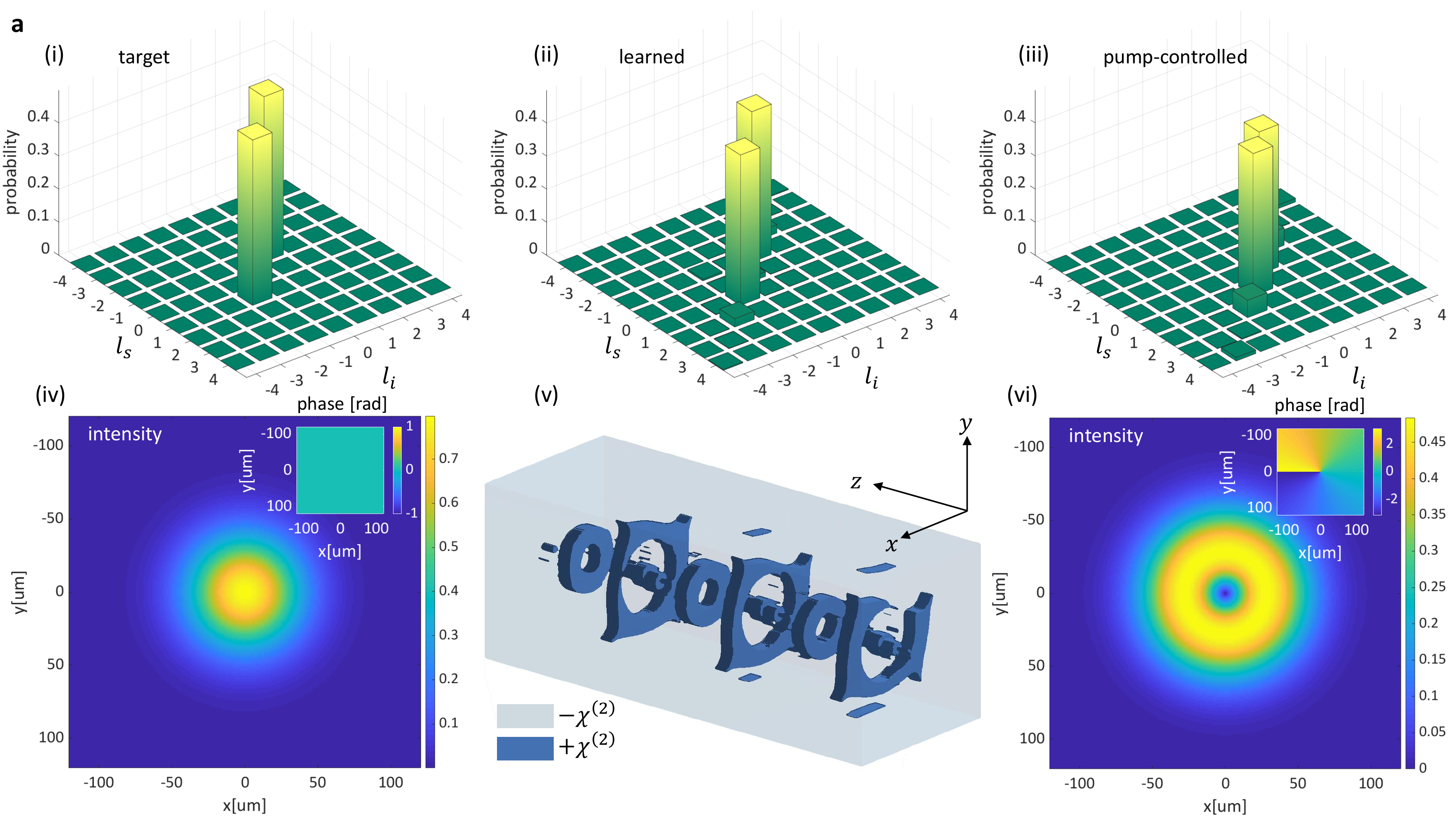}  \\
    
    \includegraphics[width=0.95\linewidth]{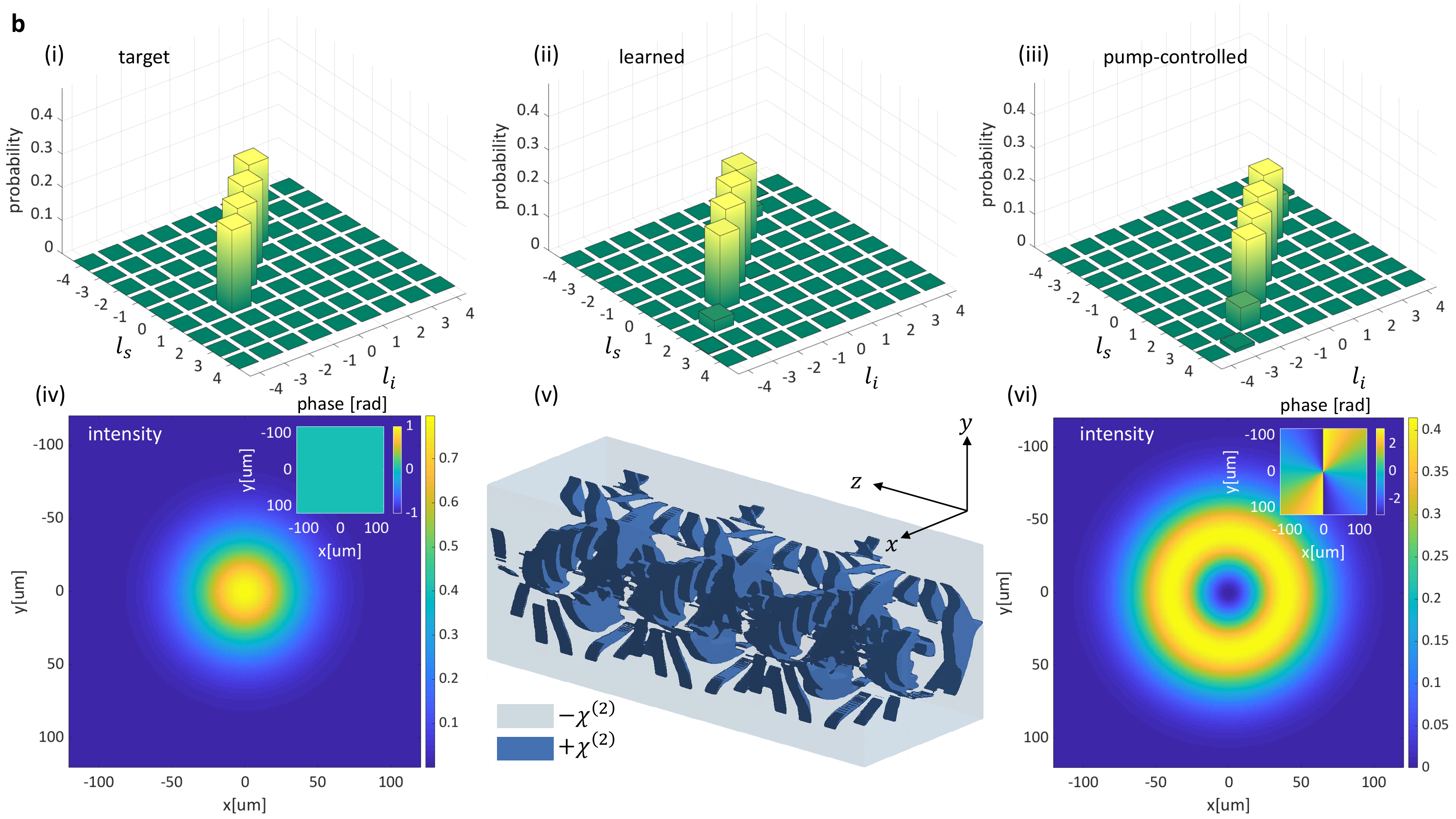}
\end{tabular}
 \caption{ \small Inverse design and all-optical coherent control over quantum correlations of SPDC photons: maximally-entangled two-photon states in the LG basis. \textbf{a}. Shaped correlations corresponding to the qubit state $\ket{\psi}=(\ket{1,-1}+\exp(i\phi)\ket{-1,1})/\sqrt{2}$. (i) shows the target coincidence probability. (ii) shows the learned coincidence probability, for an initial Gaussian pump (iv) and the learned 3D NLPC structure (v). In (v), 3 successive unit cells are shown (the z-axis is scaled-up by a factor of 20). All-optical control over the coincidence probability is demonstrated using a $\mathrm{LG_{01}}$ pump mode (vi), with the same learned crystal -- giving quantum correlations that correspond to a new qubit state, $\ket{\psi}=(\ket{0,1}+\exp(i\phi)\ket{1,0})/\sqrt{2}$ (iii). \textbf{b}. Shaped correlations corresponding to the ququart state $\ket{\psi}=(\ket{-2,1}+\exp(i\phi_1)\ket{0,-1}+\exp(i\phi_2)\ket{-1,0}+\exp(i\phi_3)\ket{1,-2})/\sqrt{4}$. (i) to (v) as in \textbf{a}. All-optical control over the coincidence probability is demonstrated using a $\mathrm{LG_{02}}$ pump mode (vi), with the same learned crystal -- giving quantum correlations that correspond to a different ququart state, residing on the $l_i + l_s = +1$ diagonal, $\ket{\psi}=(\ket{2,-1}+\exp(i\phi_1)\ket{0,1}+\exp(i\phi_2)\ket{1,0}+\exp(i\phi_3)\ket{-1,2})/\sqrt{4}$ (iii). $l_{s/i}$ are the LG modes' azimuthal integer quantum numbers for the signal/idler, respectively.}
 \label{fig:lg1}
\end{figure*}

\begin{figure*}[!ht]
\centering
\begin{tabular}{c}
    
    \includegraphics[width=0.85\linewidth]{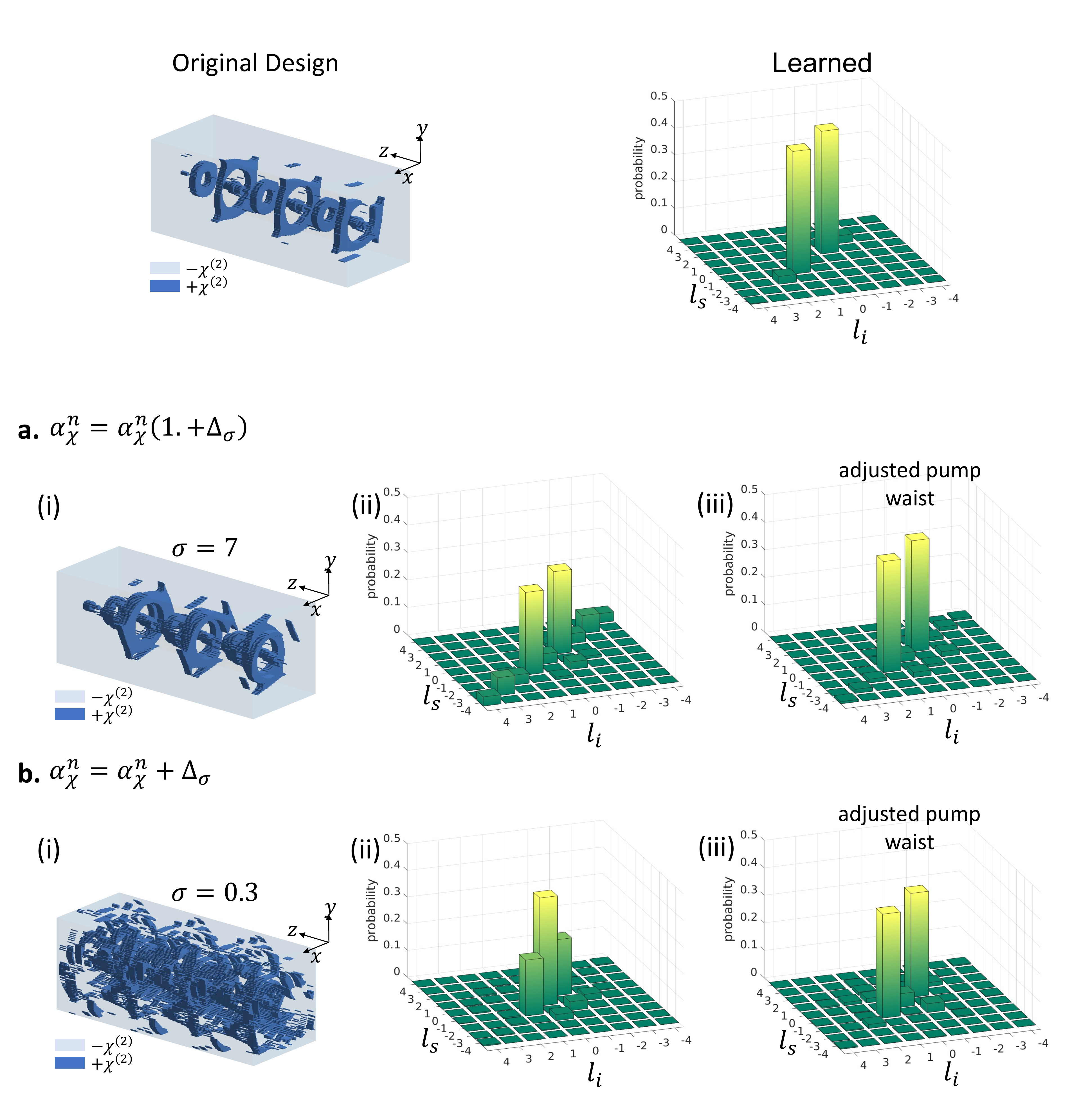}
    
\end{tabular}
 \caption{ \small Tolerance of generated coincidence rate counts of maximally-entangled two-photon qubit $\ket{\psi}=(\ket{1,-1}+\exp(i\phi)\ket{-1,1})/\sqrt{2}$, under imperfect 3D \textcolor{black}{crystal} structure. a-b(i) show two noisy versions of the optimal crystal structure (Fig. \ref{fig:lg1}.a(v)) for generating the coincidence rate counts of the desired qubit state. In a-b(i), 3 successive unit cells are shown (the z-axis is scaled-up by a factor of 20). a-b(ii) show the impaired coincidence rate counts, compare to the original setup (Fig. \ref{fig:lg1}.a(ii)). a-b(iii) show the recovered coincidence rate counts, after varying the Gaussian pump waist (Fig. \ref{fig:lg1}.a(iv)).}
 \label{fig:suppTolerance}
\end{figure*}

We now take into account a case of crystal imperfection in order to assess the tolerance of the designed crystal under fabrication errors. To do this, we first let our algorithm find the optimal spatial modes of the crystal \textcolor{black}{structure} for generating the quantum correlations of the desired quantum state with a fixed pump. After the learning phase, we deliberately add errors to the crystal structure (which mimics  crystal fabrication imperfections) and examine how does the desired quantum state is affected. We consider adding errors to the crystal coefficients in two ways (based on Eq. 8):
\begin{enumerate}[label=(\alph*)]
    \item $\alpha_\chi^n = \alpha_\chi^n (1. + \Delta_\sigma)$
    \item $\alpha_\chi^n = \alpha_\chi^n + \Delta_\sigma$
\end{enumerate}
We assume that the errors are normally distributed, i.e. $\Delta_\sigma\sim \mathcal{N}(0,\sigma^{2})$. In the first approach, there is a relative effect of the error on the amplitude of the coefficients. Although, the coefficients will always remain in the same subspace of the basis functions. In the second approach, we are no longer limited to the original subspace, but the additive noise is not correlated with the amplitude of the coefficients anymore. We present the results on the optimal \textcolor{black}{3D NLPC structures} with a constant Gaussian pump beam, for generating the desired coincidence rate counts of maximally-entangled two-photon qubit $\ket{\psi}=(\ket{1,-1}+\exp(i\phi)\ket{-1,1})/\sqrt{2}$ quantum state. Figs. \ref{fig:suppTolerance}a-b(i) present the imperfect 3D crystal design of the original design (Fig. \ref{fig:lg1}.a(v)), for the two discussed approaches. Noise was added to the coefficients until the coincidence rate counts of maximally-entangled two-photon qubit were significantly impaired relative to the original design (Fig. \ref{fig:lg1}.a(ii)), as can be seen in Figs. \ref{fig:suppTolerance}a-b(ii). At this point, we maintained the imperfect 3D \textcolor{black}{crystal} structure and tested if we can nearly recover the original system results, by modifying the Gaussian pump only (Fig. \ref{fig:lg1}.a(iv)). As can be seen in \ref{fig:suppTolerance}a-b(iii), the pump waist optimization nicely overcomes the fabrication errors, indicating the tolerance of the formed crystal. In other words, the fabrication errors diverge the model from the optimal minimum for generating the desired quantum state, but since the model was in global minimum rather than local minimum a slight variation in a different parameter of the system (pump waist) diverges the system back.

\section{Figures} \label{sec:figures}

\begin{figure*}[]
\centering
\begin{tabular}{c}
    \includegraphics[width=0.75\linewidth]{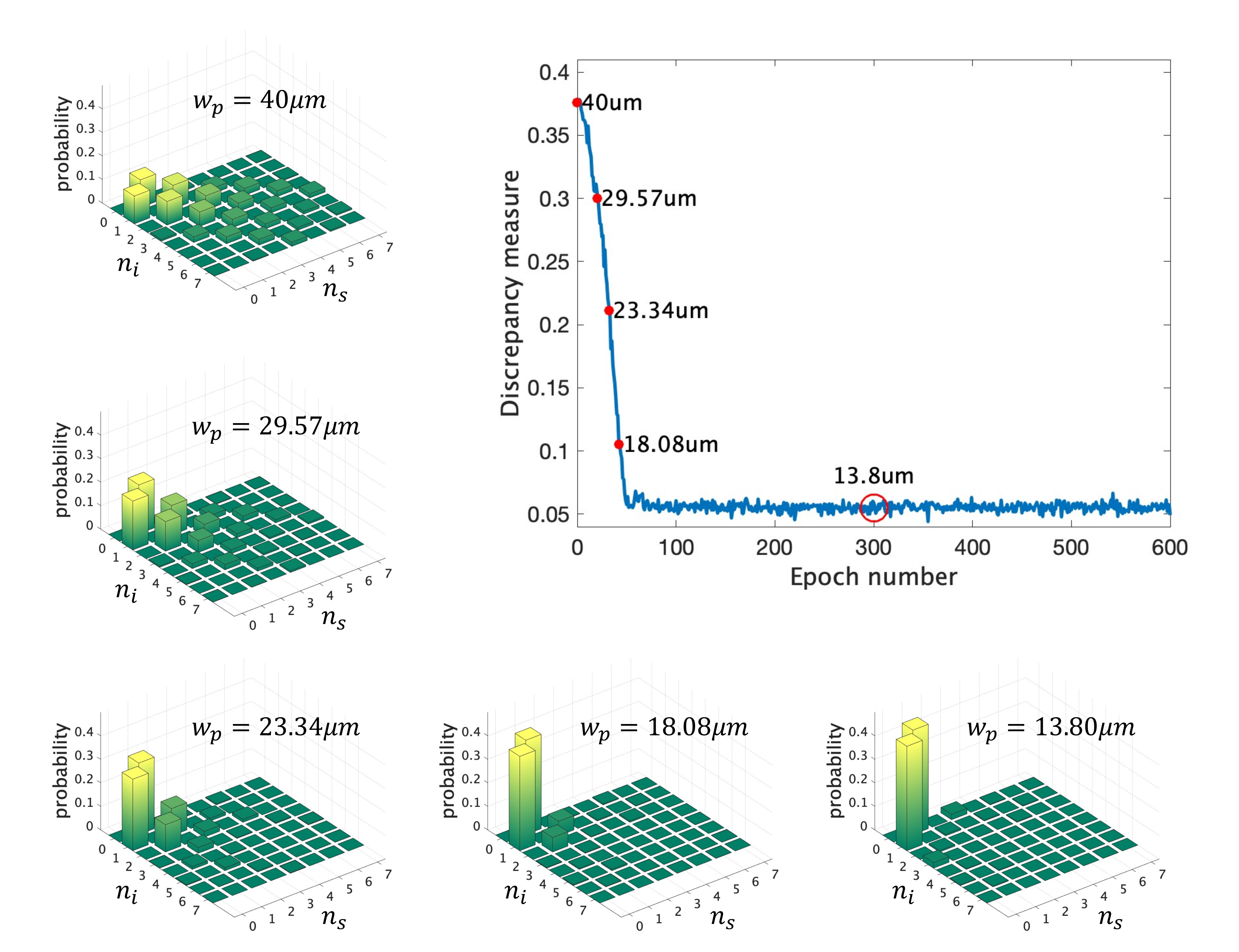}
\end{tabular}
 \caption{ \small Model validation against experimental results reported by Kovlakov et al. \citep{kovlakov2017spatial} for shaped correlations corresponding to the Bell state $(\ket{0,1}+\exp(i\phi)\ket{1,0})/\sqrt{2}$. The upper-right figure is the discrepancy measure (Eq. \ref{eq:optimizer}) between the generated coincidence rate counts and the desired one \citep{kovlakov2017spatial} vs training epoch number. The only learned physical parameter is the pump waist, and we let our algorithm find its optimal value for generating the desired quantum correlations. We sample the obtained pump waist along the discrepancy curve (red dots and insets) to see the evolution of the generated coincidence count rates under the optimized pump waist. At convergence, the algorithm obtains the correct pump waist value of $w_p=\sqrt{L/k_p}\approx13.8\mu m$ for $L=5mm$ for generating a pure HG Bell state.  $n_{s/i}$ are the HG modes' ‘X’ axis integer quantum numbers for the signal/idler, respectively.}
 \label{fig:Kovlakov_PRL}
\end{figure*}

\begin{figure}[h!]
    \centering
     $w_p=24\; \mu m$ \hspace{0.06\linewidth} $w_p=106\; \mu m$ \hspace{0.06\linewidth} $w_p=145\; \mu m$
    \includegraphics[width=0.65\linewidth]{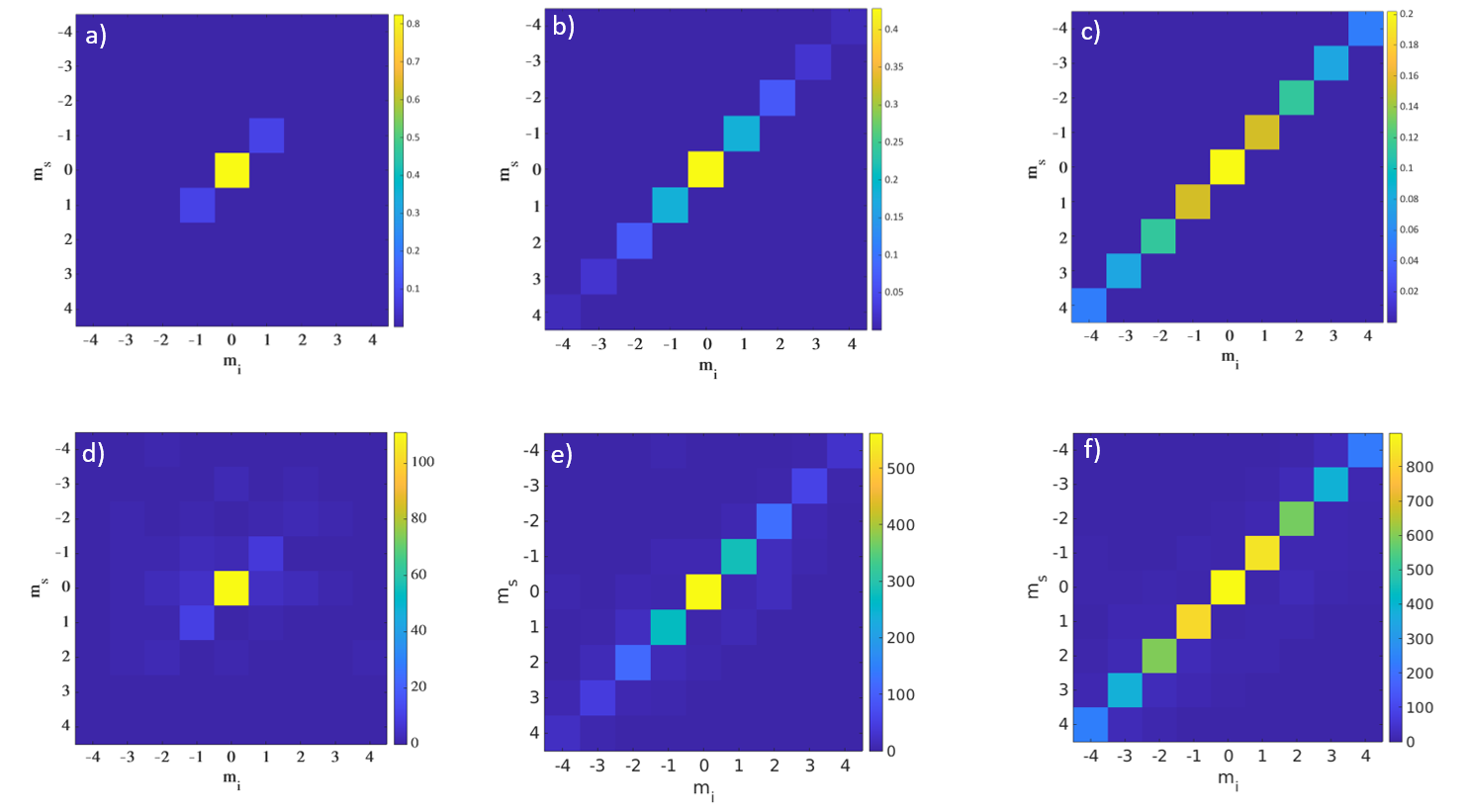}
    \caption{ \small Variation of the OAM spectrum of two photon state under different beam waists of Gaussian pump ($w_p$). Bottom row shows experimentally recorded OAM coincidence probability of signal and idler for different pump waist of Gaussian mode and top row is corresponding theoretical results.  $m_{s/i}$ are the LG modes' azimuthal integer quantum numbers for the signal/idler, respectively.}
    \label{fig:exp_oam_spect1}
\end{figure}

\begin{figure}[h!]
    \centering
    $w=2.18\; \mu m$ \hspace{0.06\linewidth} $w=3.27\; \mu m$ \hspace{0.06\linewidth} $w=6.54\; \mu m$
    \includegraphics[width=0.65\linewidth]{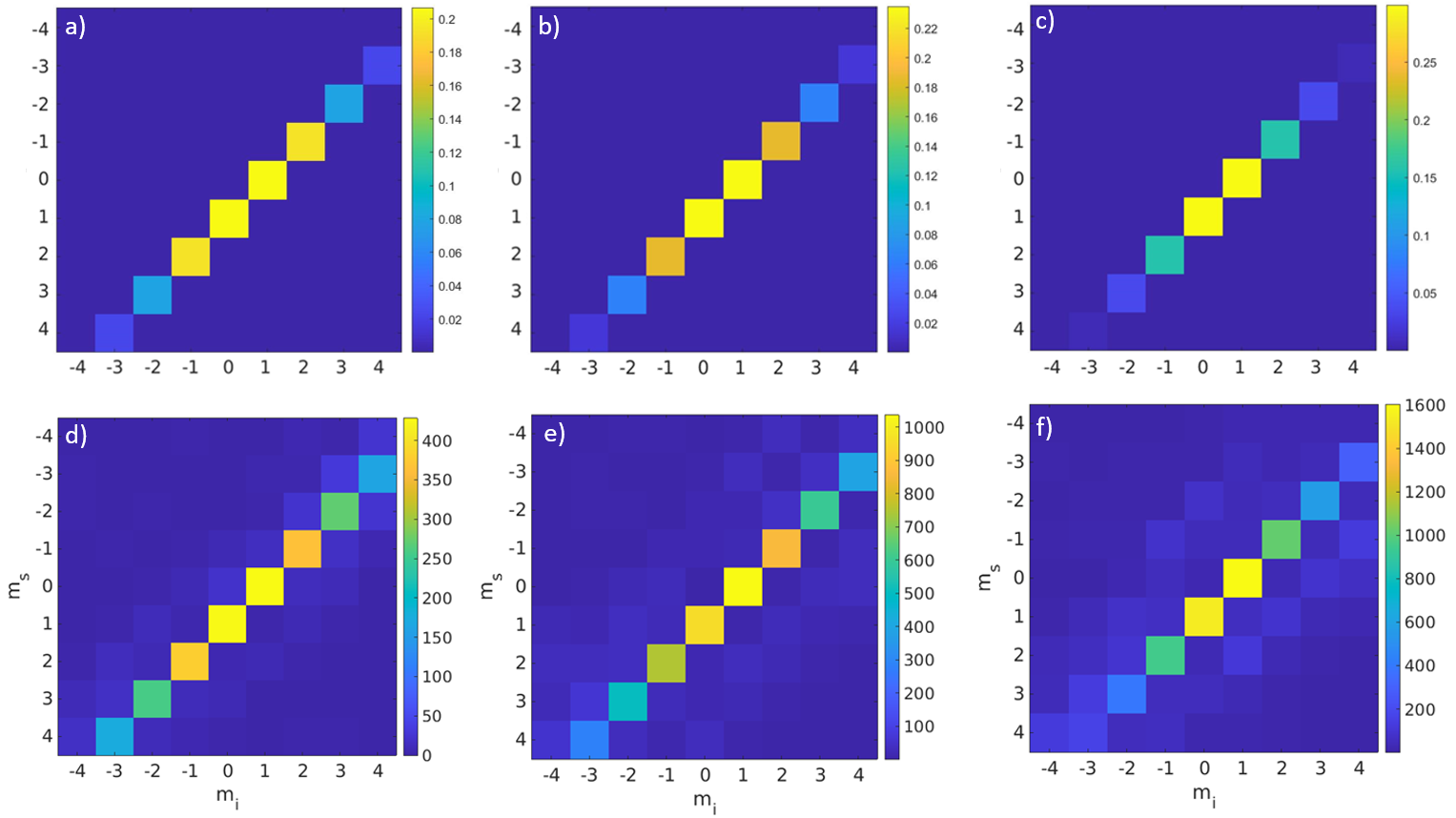}
    \caption{ \small OAM spectrum manipulated by coupling efficiency at the detection stage, for $\mathrm{LG_{01}}$ mode pump under various beam size of signal/idler ($w$) at fiber position. Mode field diameter (MFD) of SMF is $5\; \mu m$.  Top row shows the theoretical results and bottom row shows experimentally recorded data of the coincidence probability. The coupling efficiency depends on the ratio between the beam waist radius of the SMF and the size of signal/idler waist at the fiber position.  $m_{s/i}$ are the LG modes' azimuthal integer quantum numbers for the signal/idler, respectively.}
    \label{fig:exp_oam_spect2}
\end{figure}

\begin{figure}
\centering
\begin{subfigure}{.5\textwidth}
  \centering
  \includegraphics[width=.9\linewidth]{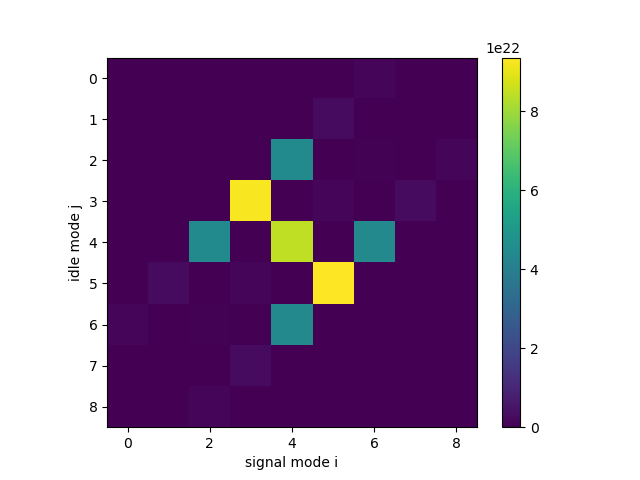}
  \caption{Numerical result}
  \label{fig:sub1}
\end{subfigure}%
\begin{subfigure}{.38\textwidth}
  \centering
  \includegraphics[width=.9\linewidth]{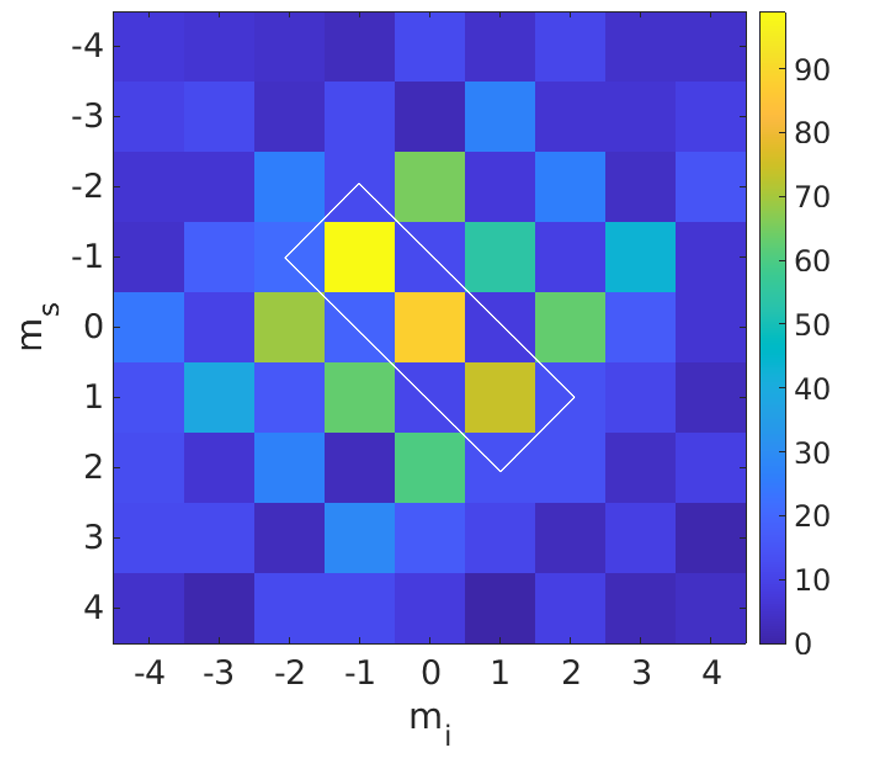}
  \caption{Experimental result}
  \label{fig:sub2}
\end{subfigure}
\caption{ \small Generation of coincidence probability of corresponding qutrit entangled state, $\ket{\psi}=(\ket{-1,1}+\exp(i\phi_1)\ket{0,0}+\exp(i\phi_2)\ket{1,-1})/\sqrt{3}$.  $m_{s/i}$ are the LG modes' azimuthal integer quantum numbers for the signal/idler, respectively.}
\label{fig:exp_qutrit}
\end{figure}

\begin{figure*}[h]
\centering
\begin{tabular}{c}
    \includegraphics[width=0.65\linewidth]{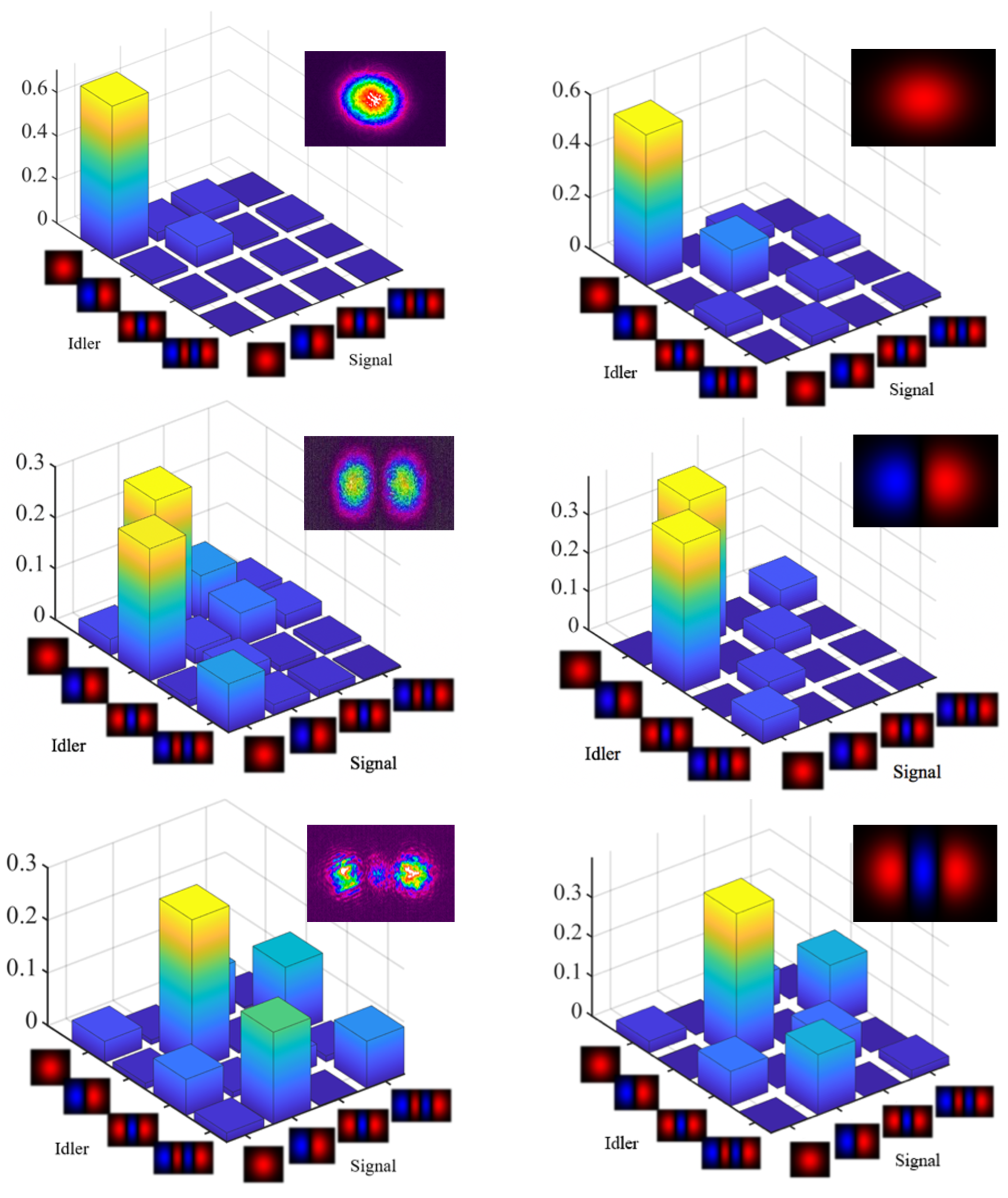}
\end{tabular}
 \caption{ \small Classical nonlinear holography and normalized coincidence detection rate for quantum nonlinear holography. Normalized coincidence counts that corresponds to the generated quantum state coefficients in the HG basis (‘X’ axis) for different NLPC structures. Left – experimental measurements. Right – corresponding simulations. Insets: Left -Second harmonic generation from the NLPC structure. Right – theoretical amplitudes of HG beam. (top) When the crystal is a regular PPKTP, only a single coefficient is nonzero. (middle) For an $\mathrm{HG_{10}}$ shaped crystal the generated state is a Bell state of the $\mathrm{HG_{00}}$ and $\mathrm{HG_{10}}$ modes. (bottom) For the short $\mathrm{HG_{20}}$ shaped crystal, the state coefficients are no longer concentrated, and exhibit even parity of the sum of the horizontal mode indices. (In courtesy of \citet{ofir_holo})}
 \label{fig:rho_exp}
\end{figure*}

\begin{figure*}[]
\centering
\begin{tabular}{c}
    \includegraphics[width=0.95\linewidth]{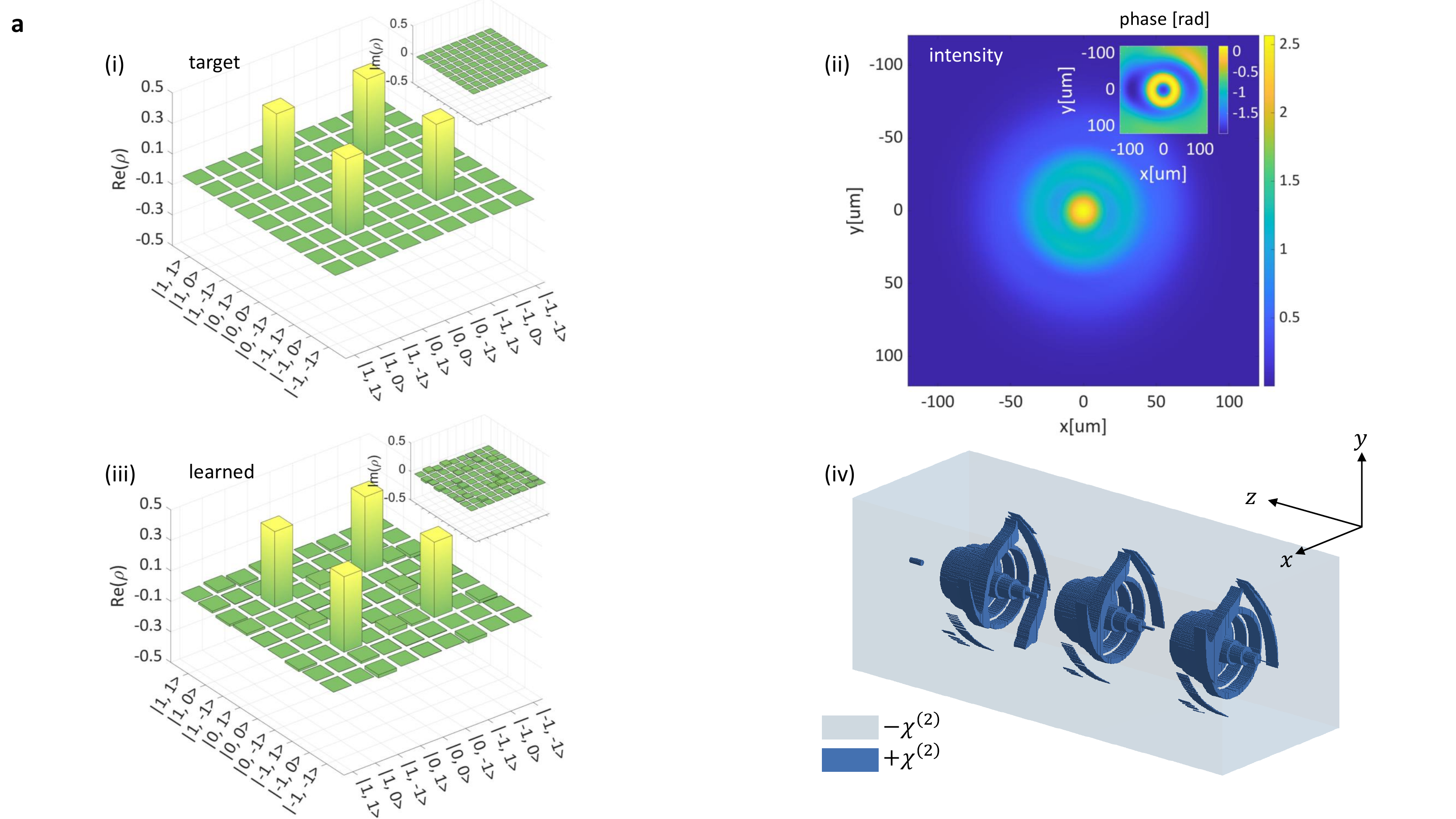}  \\
    
    \includegraphics[width=0.95\linewidth]{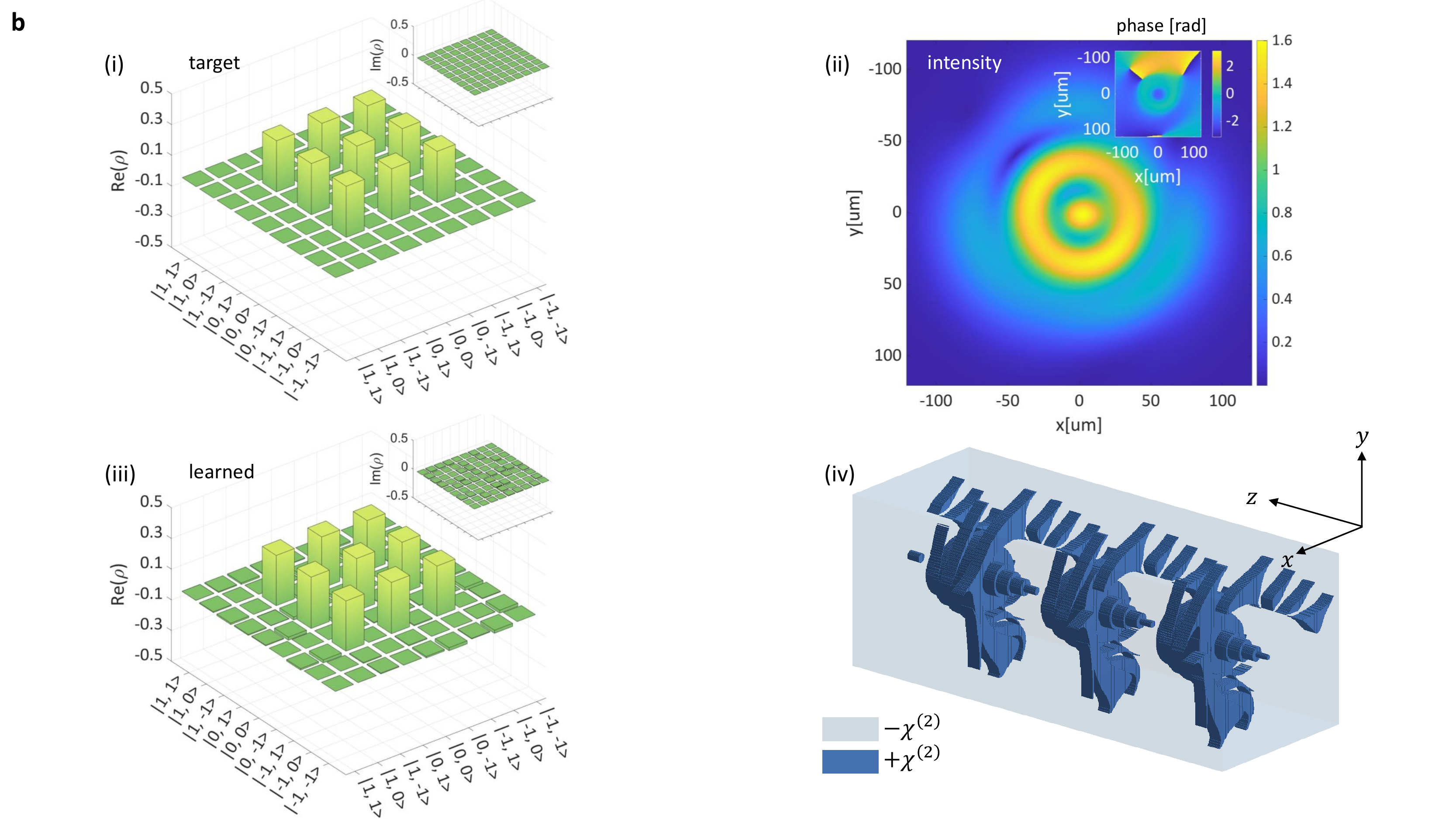}
\end{tabular}
\caption{ \small Inverse design of quantum state density matrices of SPDC photons: maximally-entangled two-photon states in the LG basis.  \textbf{a}. The qubit state $\ket{\psi}=(\ket{1,-1}+\ket{-1,1})/\sqrt{2}$. (i) and (iii) show, respectively, the target and learned states (the real part of the density matrix is shown in large, and the imaginary in small). (ii) and (iv) show the simultaneously learned complex pump beam profile and 3D NLPC structure. In (iv), 3 successive unit cells are shown (the z-axis is scaled-up by a factor of 20). \textbf{b}. The qutrit state $\ket{\psi}=(\ket{1,-1}+\ket{0,0}+\ket{-1,1})/\sqrt{3}$. (i-iv) as in \textbf{a}.}
 \label{fig:rho1}
\end{figure*}

\begin{figure*}[h]
\centering
\begin{tabular}{c}

    \includegraphics[width=0.95\linewidth]{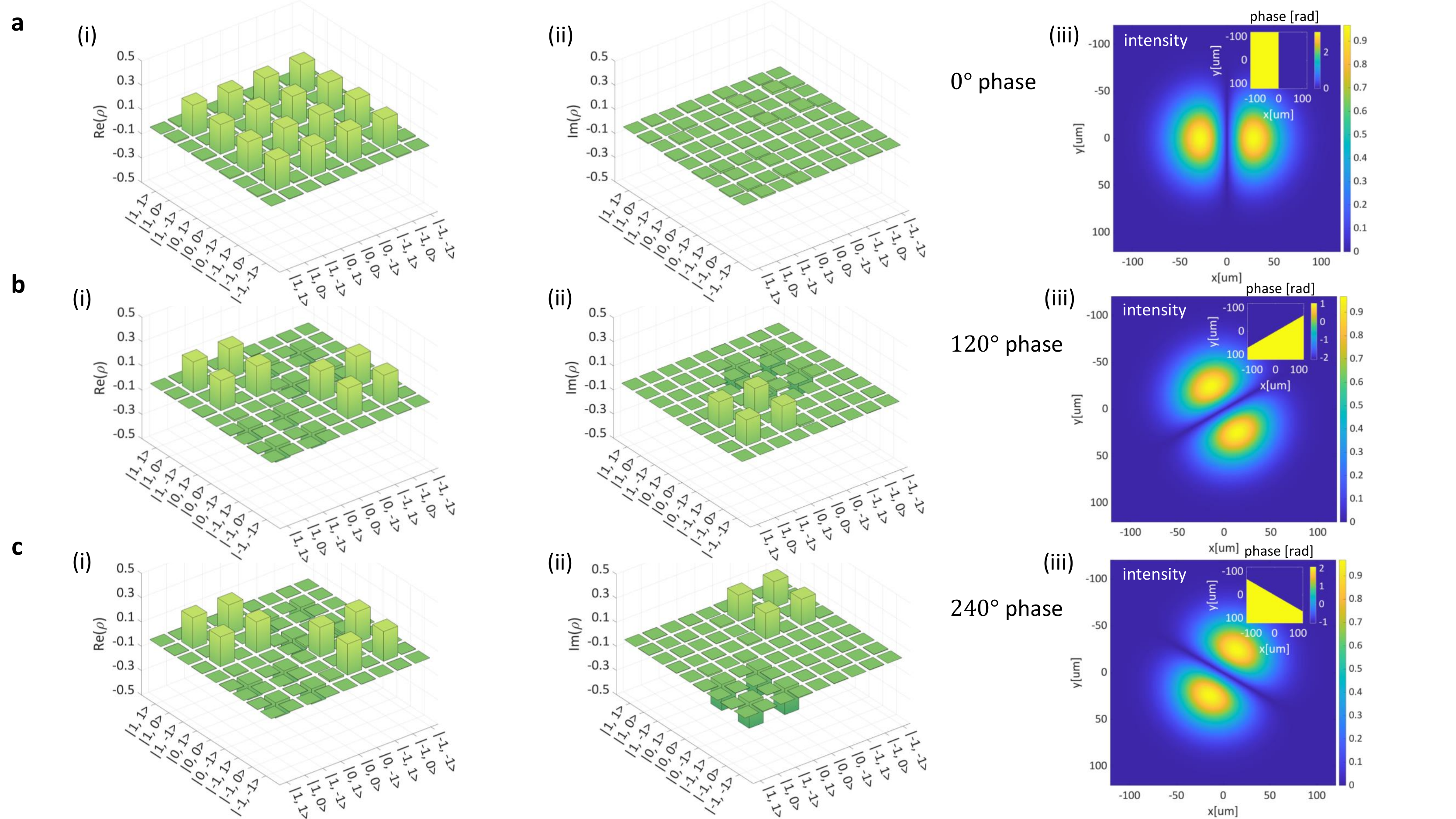}
\end{tabular}
 \caption{ \small Inverse design and all-optical coherent control over quantum state of SPDC photons: maximally-entangled two-photon ququart state in the LG basis. We use our algorithm to extract the 3D NLPC structure that generates the desired ququart state $\ket{\psi}=(\ket{-1,0}+\ket{0,-1}+\ket{1,0}+\ket{0,1})/\sqrt{4}$, using the initial constant pump profile $\mathrm{HG_{10}} = \mathrm{LG_{01}}+\mathrm{LG_{0-1}}$ 
 a(iii). The real part of generated density matrix is shown in a(i) and the imaginary part in a(ii). Next, the pump beam illuminating the learned crystal structure is rotated to actively control the generated quantum state. b(i) and (ii) show the real and imaginary parts, respectively, of generated density matrix for the rotated incident beam $\mathrm{LG_{01}}+e^{i120^{\circ}}\mathrm{LG_{0-1}}$ b(iii). c(i)-(ii) show the real and imaginary parts, respectively, of generated density matrix for the rotated incident beam $\mathrm{LG_{01}}+e^{i240^{\circ}}\mathrm{LG_{0-1}}$ c(iii). }
 \label{fig:rho2}
\end{figure*}

\end{document}